# Quantized Conductance through Surface States in High Quality Three-Dimensional Dirac Semimetal $Cd_3As_2$ Nanowire/Nanoribbon p-n Junctions

Sungjin An,[1,¶] Zhuo Bin Siu,[2,¶] Vardan Kaladzhyan,[3¶] Jens H. Bardarson,[3] Sunghun Lee,[4] Myoung-Jae Lee,[1] Kidong Park,[1] Jeunghee Park,[5] Mansoor B. A. Jalil,[2] Jungpil Seo,[6,*] and Minkyung Jung[1,7,8*]

[1] DGIST Research Institute, DGIST, Daegu 42988, Korea

[2] Electrical and Computer Engineering, National University of Singapore, Singapore 117576, Republic of Singapore

[3] Department of Physics, KTH Royal Institute of Technology, Stockholm, SE-106 91 Sweden

[4] Industrial AX Innovation Research Institute and Sensor Technology Support Center, DGIST, Daegu 42988, Korea

[5] Department of Advanced Materials Chemistry, Korea University, Sejong 30019, Korea

[6] Department of Physics and Chemistry, DGIST, Daegu 42988, Korea

[7] Department of Interdisciplinary Engineering, DGIST, Daegu 42988, Korea

[8] Institute of next generation semiconductor technology (INST), DGIST, Daegu 42988, Korea

## ABSTRACT

We report the observation of quantized conductance in high-mobility three-dimensional Dirac semimetal $Cd_3As_2$ nanowire and nanoribbon p–n junctions. By employing suspended device geometries with dual local gates, we form tunable p–n junctions and realize ballistic transport across sub-micron channel lengths. In a wide nanoribbon device with a channel width of ~ 330 nm, conductance plateaus appear at integer multiples of $2e^2/h$ in the n–n regime under high magnetic fields. Numerical simulations suggest that these features represent unresolved spin split subbands due to the smaller subband spacing in wider channels, and support the interpretation that the observed quantization may originate from surface-state-dominated conduction. In contrast, narrower nanoribbons and nanowires exhibit conductance steps of $1e^2/h$, demonstrating spin-resolved subbands likely due to enhanced confinement effects. From spin-resolved subband spectroscopy, we extract an effective Landé g-factor of ~43 for the first subband in the bulk gap, establishing these nanostructures as a prospective platform for fault-tolerant quantum electronics.

KEYWORDS: 3D Dirac semimetal, $Cd_3As_2$, conductance quantization, ballistic transport, p-n junction, surface state

¶ *S.A, Z.B.S and V.K contributed equally to this work.*

[*]Corresponding authors. Email: jseo@dgist.ac.kr, minkyung.jung@dgist.ac.kr

## INTRODUCTION

Ballistic electron transport in one-dimensional (1D) systems provides a robust framework for exploring the fundamental dynamics of quantum charge carriers.[1,2] A hallmark of ballistic transport is quantized conductance, where the conductance increases in discrete steps as successive 1D subbands are populated. Each spin-degenerate subband ideally contributes $2e^2/h$, while spin-resolved subbands yield $1e^2/h$ each. This phenomenon, first observed in quantum point contacts formed in two-dimensional electron gases (2DEGs),[1,2] has since been demonstrated in other clean 1D platforms such as carbon nanotubes[3] and semiconductor nanowires such as InAs, InSb, and PbTe.[4-13] Achieving such quantization requires transport in the ballistic regime, which demands long mean free paths, well-defined confinement, and minimal disorder. In practice, challenges such as surface scattering, structural imperfections, and non-ideal electrical contacts often obscure quantized plateaus in nanostructures.[4]

Three-dimensional (3D) Dirac semimetals (DSMs) have emerged as a new class of quantum materials ideally suited for coherent quantum devices.[14-16] In DSMs, Dirac points arise from symmetry-protected band crossings, resulting in linear energy dispersion along all three momentum directions. Among known DSMs, $Cd_3As_2$ stands out owing to its ultrahigh electron mobility ($\geq 10^5$ $cm^2/Vs$), exceptionally long transport lifetimes, and nontrivial band topology originating from band inversion between Cd-5s and As-4p states.[17-20] These properties have enabled the observation of a variety of Dirac/Weyl fermion phenomena, including negative magnetoresistance linked to the chiral anomaly, giant non-saturating magnetoresistance, and the presence of topological surface Fermi arc states.[20-24] While magnetotransport and quantum oscillations in $Cd_3As_2$ nanostructures have been extensively studied,[20-27] the direct demonstration of quantized conductance in 1D $Cd_3As_2$ channels has remained elusive. Unlike conventional semiconductor nanowires, $Cd_3As_2$ nanostructures offer an intriguing possibility, where ballistic transport occurs not only through bulk states but also through topologically protected surface channels. Theoretical studies predict that transverse confinement in nanowires discretizes these surface states into a set of 1D modes, enabling quantized transport even in the absence of strong electrostatic confinement.[28-35] Experimentally, however, resolving such quantized steps requires nanowires with minimal disorder, short transport channels compared to the phase coherence length, and reproducible low-resistance contacts, all of which pose substantial fabrication challenges.[4]

In this work, we demonstrate ballistic quantum transport in suspended $Cd_3As_2$ nanowire and nanoribbon p–n junction devices. Dual local gates allow independent control of carrier density and polarity, enabling the formation of tunable p–n junctions. Wide nanoribbons display quantized conductance plateaus at integer multiples of $2e^2/h$ in the n–n regime under high magnetic fields, characteristic of unresolved spin split subbands due to the smaller subband spacing in wider channels, while narrower nanoribbons and nanowires reveal spin-resolved $1e^2/h$ steps likely due to enhanced confinement effects. Numerical simulations suggest that these features are consistent with surface-state-mediated transport, although a direct separation of surface and bulk contributions is not established.

## EXPERIMENTAL METHODS

High-quality, single-crystalline $Cd_3As_2$ nanowires were synthesized using an optimized chemical vapor deposition (CVD) method designed to minimize defect densities. High-purity $Cd_3As_2$ powder (99%, Alfa Aesar) was utilized as the precursor inside a quartz tube reactor. To promote highly ordered one-dimensional growth, Bi catalytic nanoparticles were uniformly dispersed on a silicon substrate using an ethanol solution. Through rigorous optimization of the growth kinetics, a precise spatial temperature gradient was established: the precursor source was heated to 480°C, while the substrate—positioned exactly 8.5 cm downstream—was meticulously maintained at 350°C. A continuous flow of Argon carrier gas at 400 sccm under ambient pressure ensured a stable transport rate of the precursor vapor. A comprehensive description of the synthesis methodology and structural analysis can be found in Ref. 36 and 37. Figure 1(a) and (b) show low and high-resolution transmission electron microscopy (TEM) images, respectively, confirming the high crystalline quality of the synthesized nanowires. The high-resolution TEM image in Fig. 1(b) reveals preferential nanowire growth oriented along the [112] crystallographic direction, corresponding to the nanowire axis. A thin native oxide layer is visible on the nanowire surfaces, which was subsequently removed during device processing. The inset of Fig. 1(b) shows the calculated energy dispersion $E(k)$ near the Dirac points in the $Cd_3As_2$.[18,27] As a 3D Dirac semimetal, $Cd_3As_2$ is characterized by a band structure where the conduction and valence bands meet exclusively at discrete Dirac points within the Brillouin zone, exhibiting linear energy dispersion in all three spatial dimensions. This unique electronic structure gives rise to substantial Fermi velocities and exceptionally high mobility for three-dimensional charge carriers.[18,20]

We then fabricated suspended four-terminal devices with two recessed bottom gates to investigate the intrinsic conductance of $Cd_3As_2$ p–n junctions. Figure 1(c) and (d) present a schematic and representative scanning electron microscopy (SEM) image, respectively, of the device (Device A) and measurement configuration. The devices were fabricated on a highly doped $p^{++}$ silicon substrate covered with 300 nm of thermally grown $SiO_2$. Trenches for the bottom gates were defined via electron-beam lithography followed by anisotropic reactive ion etching with $CF_4$ and a subsequent buffered HF treatment. The resulting trenches are approximately 100 nm deep and 300 nm wide. Ti/Au (5/25 nm) bottom gates were deposited into the trenches with a spacing (pitch) of 550 nm. For Device A, a $Cd_3As_2$ nanoribbon with a width of ~330 nm and a thickness of ~80 nm was transferred across the two gates. Finally, four Ti/Au (5/120 nm) contacts, labeled electrodes 1 through 4 in Fig. 1, were patterned after in-situ Ar milling to remove the native oxide from the nanowire surface. In our previous work, oxide removal was performed using ex-situ Ar plasma, which resulted in less reliable contact quality and larger fluctuations in contact resistance.[38] To ensure consistent electrical contacts across all devices, the native oxide layer on the $Cd_3As_2$ surfaces was removed using a strictly identical in-situ Ar ion milling process immediately prior to the Ti/Au contact deposition. The milling parameters ($10^{-5}$ Torr pressure, 100 V ion energy, and 120 s) were kept constant for all fabricated devices (including Devices A, B, and C) to maintain uniform interface quality and reproducible contact resistance. The voltage probes (electrodes 2 and 3) are separated by ~920 nm. The nanoribbon segments above the bottom gates are suspended and spaced ~70 nm from the gate electrodes, allowing for efficient local gating. Four-terminal transport measurements were performed by sourcing current through electrode 1 and measuring the voltage drop between electrodes 2 and 3 using a standard lock-in technique. By comparing two-terminal and four-terminal configurations, we determined the contact resistance of Device A to be approximately 2–5 kΩ using $R_C = (R_{2t} - R_{4t})/2$, depending on the applied gate voltages (Supporting Information Fig. S1). Importantly, the four-terminal configuration directly probes the intrinsic channel conductance, eliminating contributions from contact resistance. Therefore, transport measurements in Devices A and B reflected intrinsic channel properties and are not influenced by contact resistance. All conductance data are presented without correction for series resistance. All measurements were conducted at $T \approx 2$ K.

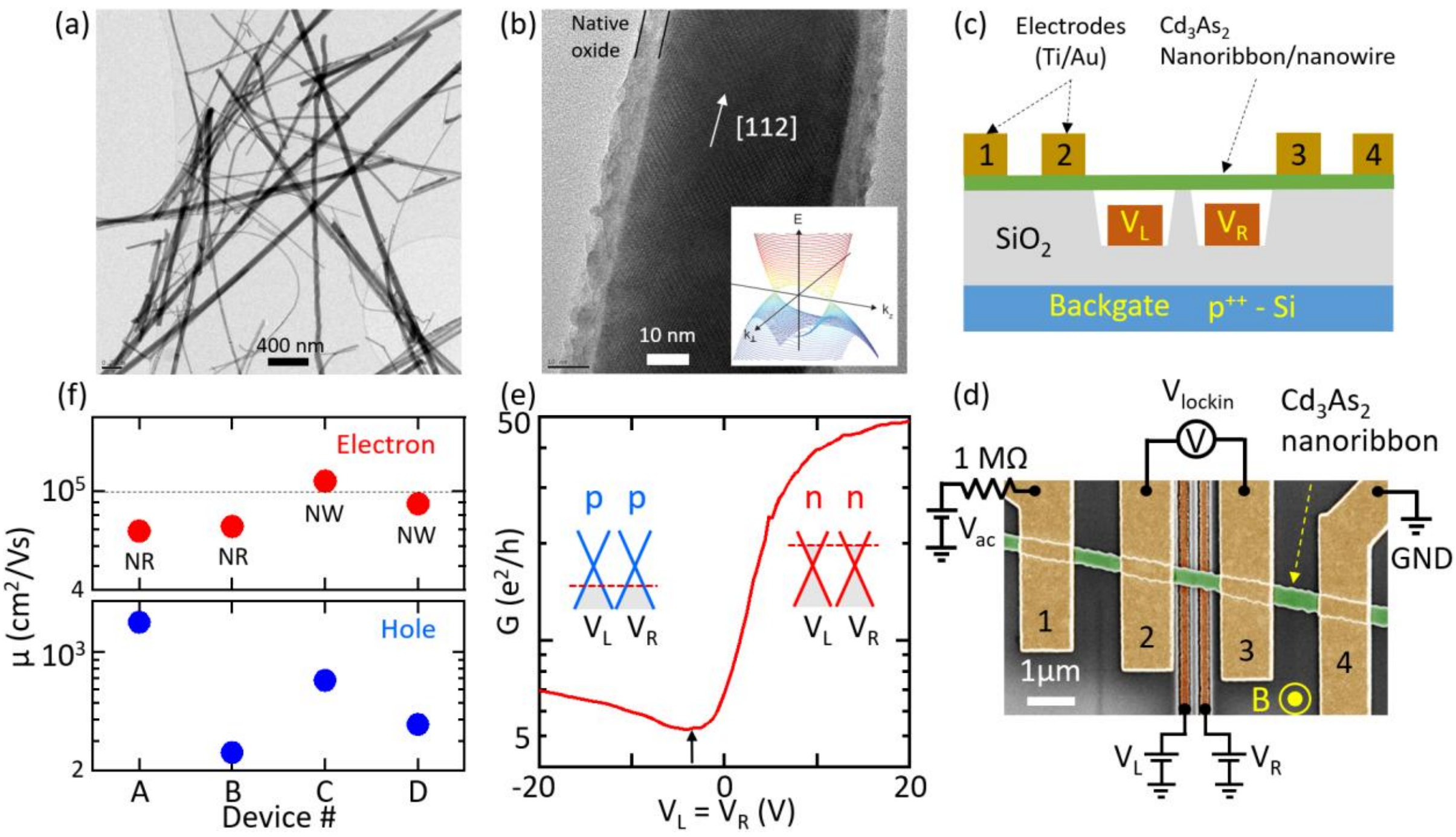

**Figure 1.** (a) Transmission electron microscopy (TEM) image of 3D Dirac semimetal $Cd_3As_2$ nanowires. (b) High-resolution TEM image of a nanowire indicates [112] growth direction. The surface of the nanowire is covered by several native oxide layers. Lower-right inset: Schematic energy dispersion $E$(k) near the Dirac nodes of $Cd_3As_2$ based on theoretical calculations. (c) Cross-sectional schematic of a suspended $Cd_3As_2$ nanowire / nanoribbon p–n junction configured for four-terminal measurement. Trenches in the $SiO_2$ were defined by anisotropic reactive ion etching (RIE) followed by wet chemical etching. Ti/Au bottom gates (5/25 nm) were deposited in the trenches. A $Cd_3As_2$ nanowire was transferred across the gates and contacted by four electrodes. (d) Representative scanning electron microscopy (SEM) image and measurement configuration of the four-terminal $Cd_3As_2$ nanoribbon device (device A). The channel length ($L$, distance between electrodes 2 and 3) and width ($W$) of the nanoribbon are ~ 920 nm and ~ 330 nm, respectively. The four-terminal transport measurements were performed using a lock-in amplifier. A perpendicular magnetic field was applied during the measurements. (e) Differential conductance as a function of the two bottom gates (with $V_L = V_R$). The conductance minimum (black arrow) indicates the location of the Dirac point in $Cd_3As_2$. (f) Extracted electron and hole field-effect mobilities from four representative devices [two nanowires (NW) and two nanoribbons (NR)].

## RESULTS AND DISCUSSION

Figure 1(e) presents the differential conductance, $G$, of Device A measured as a function of the bottom gate voltages (with $V_L = V_R$). By sweeping the two gates simultaneously, the device was tuned continuously from the p–p to the n–n unipolar transport regions. A clear conductance minimum appears near $V_L=V_R = -3.5$ V, indicating that the Fermi energy is aligned with the Dirac nodes, which further implies that Device A is weekly n-doped at zero gate voltage. The regions to the right and left of the Dirac nodes correspond to n–n and p–p doping, respectively, as illustrated by the schematic energy diagrams in the right and left insets. Electron and hole field effect mobilities extracted from $G$–$V_G$ characteristics of two nanoribbon and two nanowire devices are summarized in Fig. 1(f). The electron mobility $\mu_e$ ranges from $\sim 7 \times 10^4$ cm$^2$/V·s to $\sim 1.1 \times 10^5$ cm$^2$/V·s, substantially exceeding the values typically reported for semiconducting nanowires such as InAs and InSb.[4-11]. The highest mobility observed in nanowires reaches ~ $1.1 \times 10^5$ cm$^2$/V·s, indicating excellent electrical quality in both nanowires and nanoribbons.

Figure 2(a) presents the four-terminal conductance map of Device A measured as a function of $V_L$ and $V_R$ at $B = 0$ T. The device exhibits distinct p–n junction behavior controllable by the two recessed bottom gates, resulting in four distinct regimes (n–n, p–n, p–n and n–p) corresponding to different carrier combinations in the left and right segments. The white dashed lines indicate the locations of the Dirac nodes for the left and right nanowire segments, separating the unipolar (p–p or n–n) and bipolar (p–n or p–n) regimes. The border lines separating the four different regimes are approximately perpendicular to each other, implying that capacitive cross coupling between the two gates is very weak. The conductance in the n–n regime is markedly higher than that in the p–p regime; this is attributed to the valence band having a much larger gradient than the conduction band, as shown in the $Cd_3As_2$ band structure in the lower right inset of Fig. 1(b), which typically results in a heavier effective mass and lower mobility for holes.

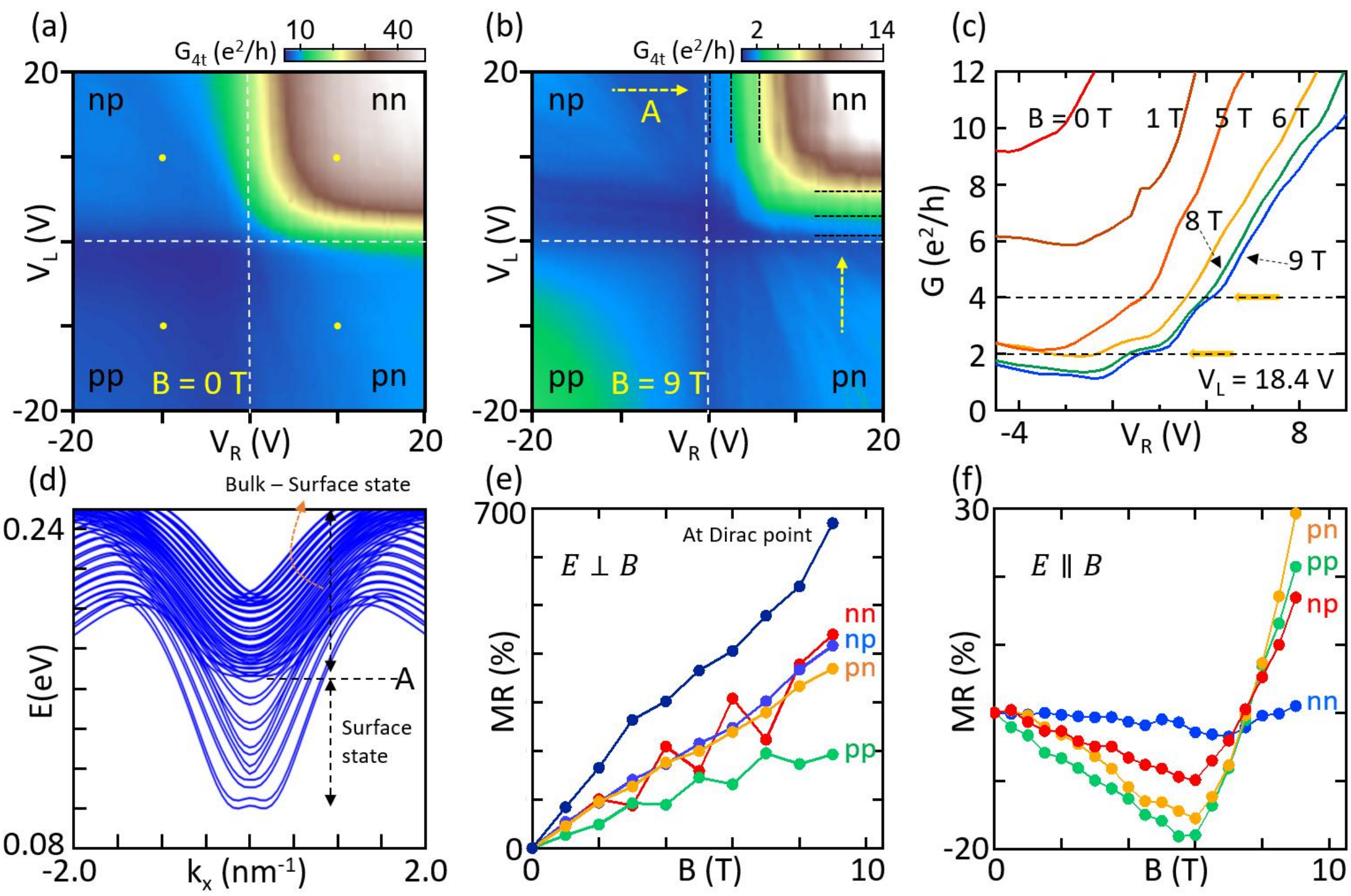

**Figure 2.** Four-terminal differential conductance map of Device A as a function of $V_L$ and $V_R$ measured at (a) $B = 0$ T and (b) $B = 9$ T. The four regions correspond to different carrier doping configurations comprising–n–n, p–p, p–n, and n–p–controlled independently by the left and right bottom gates. The white dash lines indicate the locations of the Dirac nodes. The vertical and horizontal lines are nearly orthogonal, confirming minimal capacitive cross talk between the gates. At $B = 9$ T, quantized conductance plateaus at integer multiples of $2e^2/h$ emerge in the n–n regime, as marked by the black dashed vertical and horizontal lines. (c) Line cuts along the direction A (yellow dashed arrow in (b)) as a function of $V_R$ and magnetic field at $V_L = 18.4$ V. (d) Band structure for a an infinite $Cd_3As_2$ nanowire aligned along the x axis. (e) Magnetoresistance (MR), defined as $[R(\mathrm{B}) - R(0)]/R(0) \times 100$ (%), measured in the four representative p–n junction configurations (yellow dots in (a)) and at Dirac point, with the magnetic field applied perpendicular to the electric field ($E \perp B$) up to $B = 9$ T. All regimes exhibit a large positive and non-saturating MR, reaching ~670 % at $B = 9$ T near the Dirac point. (f) MR measured with the magnetic field applied parallel to the electric field ($E \parallel B$). All doping configurations display negative MR behavior, consistent with previously reported longitudinal magnetoresistance associated with the chiral anomaly in $Cd_3As_2$.[21,22]

Figure 2(b) shows the conductance map measured at $B = 9$ T. Pronounced quantized conductance plateaus appear in the n–n regime (highlighted by black dashed lines), which are absent at $B = 0$ T. The observed conductance quantization is attributed to 1D subband formation arising from transverse confinement in the nanowire. However, at $B = 0$ T, the plateaus are not clearly resolved, likely due to backscattering and mode mixing, which can smear the subband features. From the measured mobility, we estimate a mean free path on the order of ~ 1 μm, comparable to the channel length (~ 900 nm), indicating quasi-ballistic transport. However, residual scattering remains sufficient to obscure quantization at zero field. As the magnetic field increases, the visibility of quantization is enhanced, as it can suppress backscattering and reduce mode mixing, while also introducing additional magnetic confinement.[4,6] These effects are expected to sharpen the subband structure and make the underlying quantized conductance more apparent. Figure 2(c) presents line cuts along direction A in Fig. 2(b), illustrating the conductance evolution with $V_R$ and $B$ at $V_L = 18.4$ V. Similar quantization is widely observed in semiconductor nanowires (e.g., InAs, InSb) where each spin-degenerate 1D subband contributes a quantum of $2e^2/h$ to the conductance at $B = 0$T.[6,7] Under an applied magnetic field, Zeeman splitting typically lifts this degeneracy, resulting in conductance steps of $1e^2/h$. However, in our $Cd_3As_2$ nanoribbon geometry (Device A), the observed plateaus remain at multiples of $2e^2/h$. This is likely due to the resulting Zeeman splitting ($\Delta E_Z = g^* \mu_B B$) being insufficient to overcome level broadening in the wide and thick nanoribbon under these measurement conditions. The effective Landé g-factor has been shown to be dependent on device geometry and quantum confinement in semiconductor nanostructures.[39] In Device A, the wider dimensions lead to a renormalization of the effective g-factor and reduced subband spacing that prevents the resolution of spin splitting, preserving the apparent $2e^2/h$ plateaus.

To elucidate the origin of the observed conductance quantization and determine whether transport is dominated by bulk or surface states, we perform numerical calculations of the energy band structure using a 3D Dirac semimetal Hamiltonian (Supplement note S2 in the Supporting Information).[18,27,28] These calculations incorporate both orbital and Zeeman effects, providing a comprehensive picture of the electronic structure under experimental conditions. Figure 2(d) shows the calculated band structure for a nanowire infinite along the transport direction ($x$), plotted as a function of the $k_x$, which represents a "good" quantum number in this geometry. The calculated spectrum highlights the finite-size effects of topological semimetals. Due to transverse confinement, the bulk states acquire a sizable energy gap (indicated by the horizontal dashed line A), while the topological surface modes persist within this bulk gap. Since the surface wavefunctions are confined to the nanowire perimeter whereas bulk wavefunctions extend across the full cross-section, the confinement gap for the bulk states is significantly larger than that for the surface states. Consequently, only surface state subbands exist within the bulk gap region. These form a series of discrete one-dimensional modes with small energy separations determined primarily by the Zeeman effect. To further quantify this distinction, we estimate the characteristic energy scales. For a nanoribbon with width $W$ ~ 300 nm and height $H$ ~ 80 nm, the surface-state subband spacing is determined by the perimeter $P$ ~ 760 nm, yielding $\Delta_{Surf} \sim \mathrm{h}v_F/p \approx 5\, meV$ for $v_F = 1 \times 10^6$ m/s. In contrast, the bulk confinement energy scale, set by the smallest transverse dimension, is $\Delta_{bulk} \sim \pi\hbar v_F/H \approx 26\, meV.$ The significantly larger bulk confinement gap implies the existence of an energy window in which bulk states are suppressed while surface states remain active. In this regime, transport is expected to be dominated by surface channels, consistent with the observed conductance quantization near the Dirac point.[28]

Although the applied magnetic field induces Zeeman splitting, the energy splitting between adjacent surface subbands remains smaller than the experimental energy broadening caused by thermal and disorder effects. As a result, the conductance plateaus appear at integer multiples of $2e^2/h$, consistent with unresolved spin-split modes in our device geometry and measurement temperature. More generally, the visibility of the subband structure is governed by the ratio between the subband spacing Δ and the effective broadening Γ, arising from disorder and finite quasiparticle lifetime. When $\Gamma \ll \Delta$, individual subbands can be resolved, whereas for $\Gamma \gtrsim \Delta$, the subband features become progressively smeared out. The numerical simulations presented here correspond to the clean limit, where Γ is negligible and the underlying subband structure is clearly resolved. In the experimental devices, residual disorder and scattering introduce finite broadening, which can reduce the visibility of quantized conductance, particularly at zero magnetic field.

To further characterize the electronic properties and confirm the Dirac semimetal nature of our devices, we examine the magnetoresistance (MR), defined as $[R(\mathrm{B}) - R(0)]/R(0) \times 100$ (%), in different p–n configurations under two magnetic field orientations. For the perpendicular configuration ($E \perp B$, Fig. 2(e)), all regimes exhibit a large positive MR, with the maximum MR reaching 670 % around the Dirac point at $B = 9$ T. Such giant, non-saturating positive MR is a hallmark of three-dimensional Dirac semimetals arising from the field-induced suppression of the unusual backscattering protection present at zero field.[20,23,24] In $Cd_3As_2$, ultrahigh mobilities and long transport lifetime are enabled by a mechanism that strongly suppresses large-angle scattering processes. An applied magnetic field breaks this protection-likely by modifying the Dirac Fermi surface-leading to a dramatic increase in resistivity. In contrast, for the parallel configuration ($E \parallel B$, Fig. 2(f)), all doping configurations exhibit a clear negative MR. This negative MR is consistent with the chiral anomaly effect characteristic of Dirac and Weyl semimetals where parallel electric and magnetic fields create an imbalance in the population of Weyl nodes with opposite chiralities, resulting in enhanced conductivity.[21,22] The magnitudes of both positive and negative MR vary between different doping configurations, which we attribute to variations in junction resistance and carrier transmission probabilities across the p–n interfaces. The observation of large positive and negative magnetoresistance further reflects the distinctive Dirac semimetal nature of $Cd_3As_2$, including the suppression of backscattering and chiral anomaly effects.

We next investigate the conductance quantization in a $Cd_3As_2$ nanowire (Device B), which is weakly p-doped at zero gate voltage, with a diameter of ~100 nm, as shown in the SEM image in Fig. 3(a).[38] Using the four-terminal configuration, we measured the differential conductance as a function of $V_L$ and $V_R$ at $B$ = 0 T (Fig. 3(b)) and 9 T (Fig. 3(c)). In contrast to Device A (Fig. 2), this device exhibited more pronounced capacitive cross-talk between bottom gates, as indicated by the sloped yellow dashed lines. At $B$ = 9 T (Fig. 3(c)), conductance plateaus appeared in the n–n regime at integer multiples of $1e^2/h$ rather than $2e^2/h$, as marked by the black dashed lines (regions A and B). Figure 3(d) presents a line cut along direction C (yellow dashed arrow in Fig. 3(c)) at $V_L$ = 13.5 V, clearly revealing plateaus at $G$ = $1e^2/h$ and $2e^2/h$, corresponding to regions A and B, respectively. This behavior contrasts with Device A, where the first well-defined plateaus appeared at multiples of $2e^2/h$.

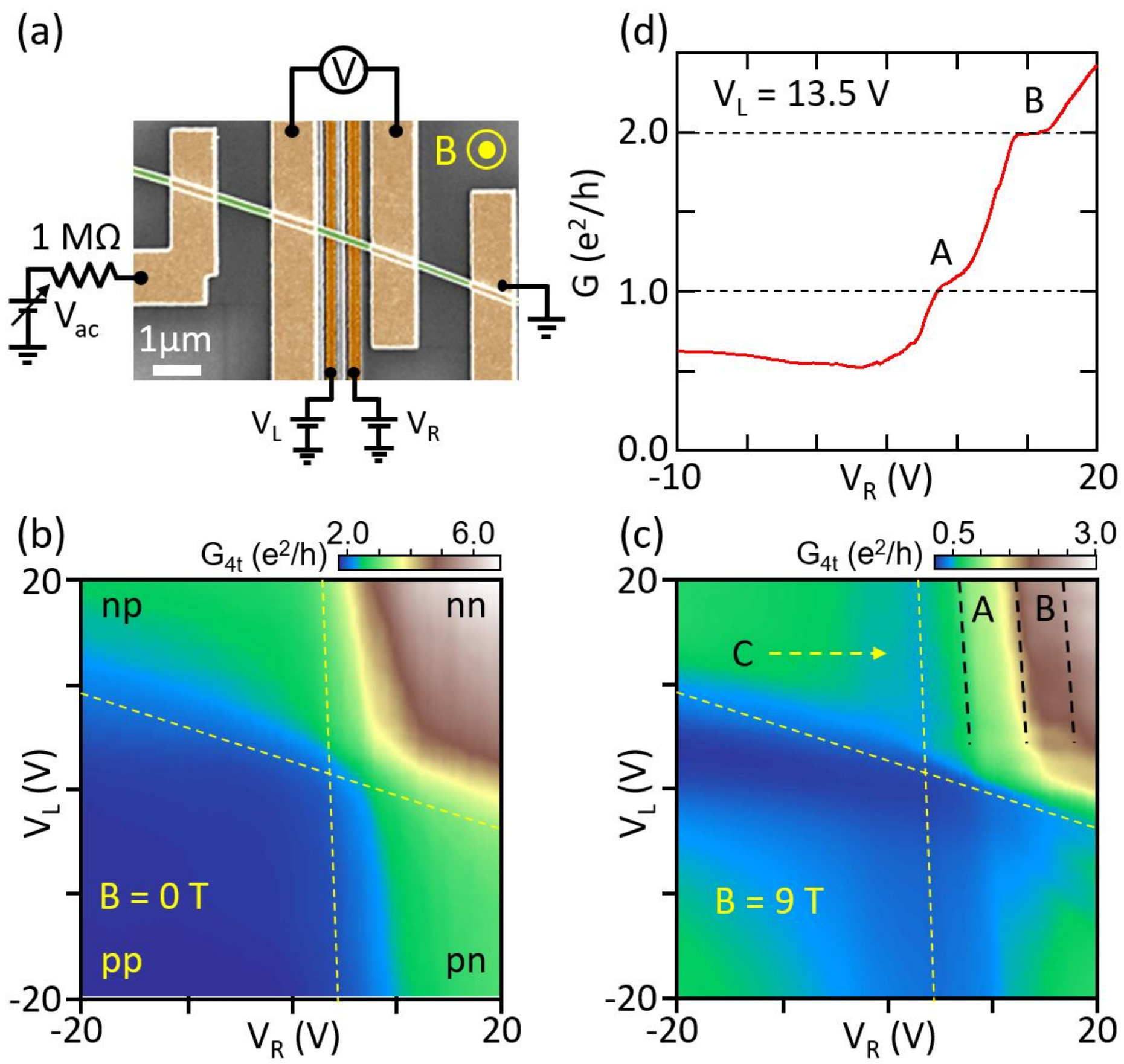

**Figure 3.** (a) Four-terminal $Cd_3As_2$ nanowire device with recessed bottom gates (Device B). The channel length ($L$) and diameter ($D$) of the nanowire for Device B are ~ 900 nm and ~ 100 nm, respectively. Four-terminal differential conductance map as a function of $V_L$ and $V_R$ at (b) $B$ = 0 T and (c) $B$ = 9 T. At $B$ = 9 T, conductance plateaus appeared in the n–n regime at the multiples of $1e^2/h$ indicated by the black dashed lines. (d) Line trace taken along the C direction (yellow dashed arrow, $V_L$ = 13.5 V) in (c). Weak conductance plateaus are observed at 1 and $2e^2/h$, corresponding to the A and B regimes in (c).

The emergence of $1e^2/h$ steps in Device B can be attributed to stronger confinement effects in the narrower nanowire geometry, which enables the resolution of spin-split subbands. In one-dimensional ballistic channels, each spin-resolved subband ideally contributes $1e^2/h$ to the conductance. Strong transverse confinement, which is enhanced by a high magnetic field, can lift the spin degeneracy and resolve the first spin-polarized mode. This behavior parallels reports on InSb and InAs nanowires, where increasing $B$ suppressed backscattering, enhanced subband separation, and revealed spin-resolved conductance plateaus.[4-11] The difference between Devices A and B can therefore be understood as a combined effect of geometry and mode spacing. The smaller diameter of Device B results in a large subband energy spacing, making the Zeeman-split subbands resolvable at $B$ = 9 T. In contrast, the wider nanoribbon in Device A exhibits smaller subband spacings, so spin splitting is unresolved, and the plateaus remain at multiples of $2e^2/h$.

We further confirm quantized conductance in a narrower and thinner $Cd_3As_2$ nanoribbon (Device C), which is weakly p-doped at zero gate voltage, using a two-terminal measurement configuration, which yields a more pronounced conductance plateau. The nanoribbon in Device C has a width of ~ 200 nm and thickness of ~ 50 nm, as shown in the SEM image in the inset of Fig. 4(a). In this device, only the left gate ($V_L$, highlighted in orange) is operational. Figure 4(a) displays the differential conductance as a function of $V_L$ at $B$ = 8 T, plotted in units of $e^2/h$. The Dirac point appears near $V_L$ = 7.7 V. As $V_L$ increases, the conductance exhibits a pronounced plateau near $0.94e^2/h$. The deviation from the ideal $1e^2/h$ value is likely due to a finite series resistance, dominated by contact resistance (~ a few kΩ), which is typical for such nanoribbon/nanowire devices. After accounting for this series resistance, the plateau height closely approaches the theoretical quantized value ($1e^2/h$), signifying ballistic transport through a single spin-resolved 1D subband. Figure 4(b) displays the evolution of conductance with magnetic field. At $B$ = 0 T, the plateau is absent, likely due to strong backscattering and mode mixing. As $B$ increases, magnetic confinement and suppression of backscattering promote ballistic transport, revealing the first spin-resolved plateau at $1e^2/h$.

More generally, the characteristic subband spacing is expected to scale inversely with the relevant transverse dimension. For the nanoribbon (Device C), the estimated surface-state subband spacing is on the order of 5–8 meV, while for the thicker nanowire devices (Devices A and B), the larger transverse dimensions lead to smaller subband spacing. This scaling is consistent with the experimental observation that conductance quantization is more clearly resolved in the narrower device, while spin splitting remains unresolved in the wider devices due to smaller subband separation relative to the broadening.

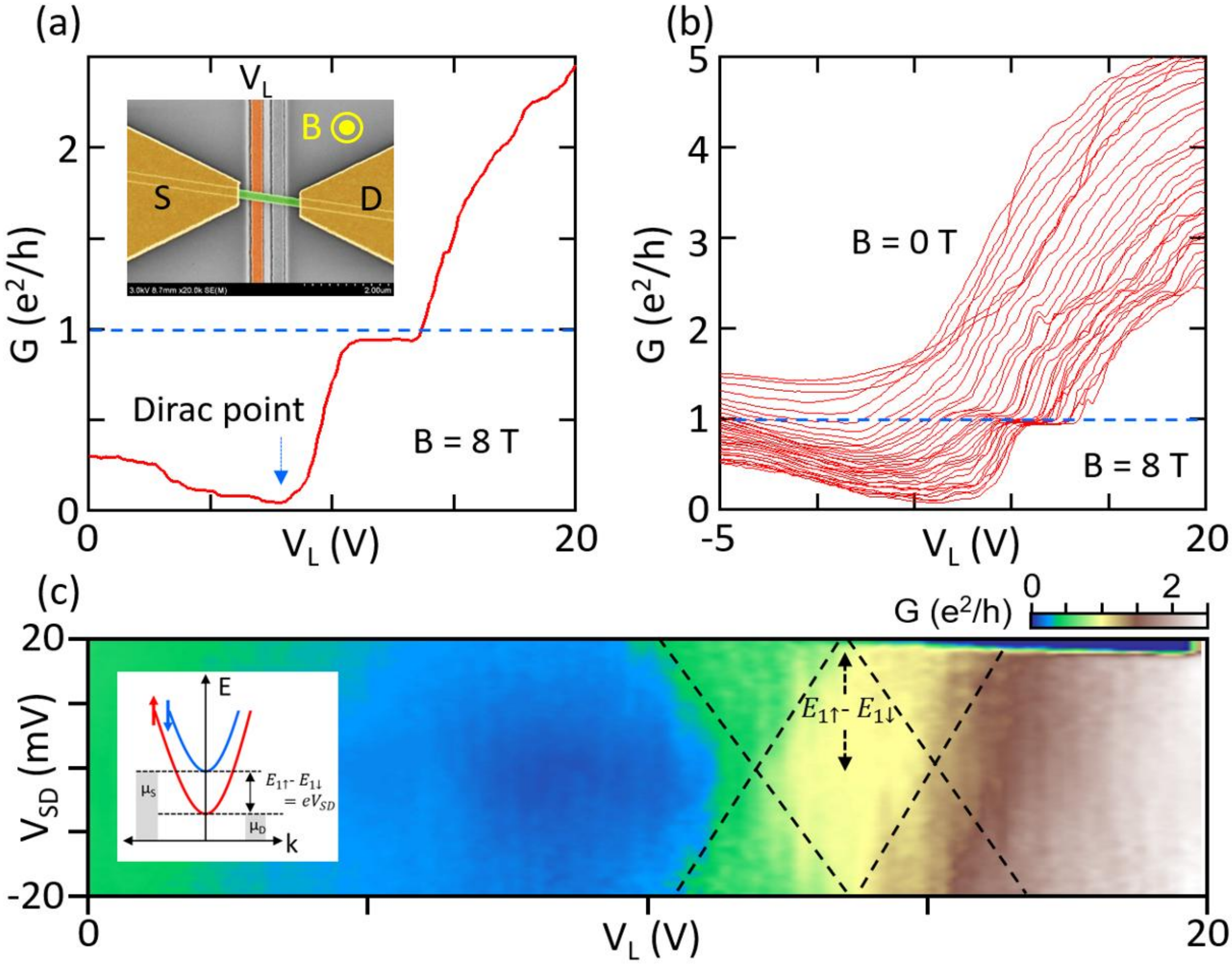

**Figure 4.** (a) Conductance measured in Device C at $B$ = 8 T. The inset show a SEM image of Device C. The channel width and thickness are ~ 200 nm and ~ 50 nm, respectively. In this device, the left gate ($V_L$) is operational. (b) Conductance measured as a function of magnetic field. (c) 2D conductance map of the differential conductance as a function of $V_{SD}$ and $V_L$ at $B$ = 8 T. Black dotted lines surrounding a diamond shaped region, indicating the edge of the first quantized conductance plateau, are drawn as guide to the eye. Inset: energy spectra showing the lowest two spin-split subbands at $V_{SD} \neq 0$ mV. The subband spacing is equivalent to the source-drain voltage ($\Delta E_{suband} = E_{1\uparrow} - E_{1\downarrow} = eV_{SD}$).

To characterize the subband structure, we measured the differential conductance at $B$ = 8 T as a function of $V_{SD}$ and $V_L$ (Fig. 4(c)). The resulting color plot reveals a diamond-shaped region of constant conductance, outlined by black dashed lines. In the center of the diamond, a prolonged plateau at 0.94e$^2$/h appears near $V_{SD}$ = 0 V, corresponding to a chemical potential ($\mu_S = \mu_D$) lying between the first spin-split subbands $E_{1\downarrow}$ and $E_{1\uparrow}$. At the diamond tips where the dashed lines intersect, the applied bias matches the energy spacing between these subbands, $\Delta E_{suband} = E_{1\uparrow} - E_{1\downarrow} = eV_{SD}$. From the Zeeman splitting $\Delta E_{suband} \approx$ 20 meV, we extract the effective Landé g-factor of the first subband using the Zeeman relation, $\Delta E_Z = g^* \mu_B B$, where $\mu_B$ is the Bohr magneton and $B$ = 8 T. This yields $g^* \approx 43$, in close agreement with the value estimated from previous studies.[24,27] In our modeling, both orbital and Zeeman effects of the magnetic field are included. Therefore, the evolution of the subband structure under magnetic field reflects their combined influence. The extracted $g^*$ should thus be regarded as an effective Landé factor that may include both spin and orbital contributions.

## CONCLUSION

In summary, we have demonstrated ballistic quantum transport in high-mobility 3D Dirac semimetal $Cd_3As_2$ nanowire and nanoribbon p–n junction devices. The suspended dual-gated geometry provides precise electrostatic control, enabling the systematic investigation of conductance. In wide nanoribbon devices, quantized plateaus at integer multiples of $2e^2/h$ were observed in the n–n regime under high magnetic fields, consistent with transport through unresolved spin split 1D subbands. In narrower nanowire and nanoribbon devices, we resolved conductance steps at $1e^2/h$, indicating spin-resolved subband transport enabled by stronger geometric confinement and enhanced subband separation under magnetic field. Subband spectroscopy, performed by measuring conductance quantization as a function of source-drain voltage and gate voltage, allowed us to extract an effective Landé g-factor of ~43 and a subband spacing of ~20 meV. Numerical simulations suggest that the observed quantization is consistent with a dominant contribution from surface states, although a direct experimental distinction from bulk transport remains an open question. Our findings establish $Cd_3As_2$ nanowires as a promising platform for quantum electronics, in which high mobility and robust surface channels can be utilized for next-generation spintronics and fault-tolerant quantum processing.

## Supporting Information

Contact resistance of Device A (Supplement Notes S1, Figures S1), Modelling and numerical simulations (Supplement Notes S2) (PDF)

**ACKNOWLEDGEMENTS**

This work is supported by the Mid-Career Researcher Program of NRF (RS-2023-NR076585 and RS-2023-00278511) and the DGIST R&D Program of the Ministry of Science, ICT, and Future Planning (26-ET-02, 25-SENS2-09, 26-SR-02). This work was supported by Institute for Information & communications Technology Promotion (IITP) (2021-0-01511), and NRF (RS-2023-00269616, RS-2023-00258732, RS-2024-00402302, RS-2025-25463923, RS-2023-00285353, RS-2025-02317602, RS-2025-02315685, RS-2025-16070349) grants funded by the Korean government (MSIT & MOE). Z.B.S. and M.B.A.J. are supported by a Singapore MOE Tier-I FRC grant (NUS grants A-8002026-00-00). This work received funding from the Swedish Research Council (VR) through Grant No. 2020-00214 and the Knut and Alice Wallenberg Foundation (KAW) via the project Dynamic Quantum Matter (2019.0068).

**AUTHOR CONTRIBUTIONS**

S.A. and K.P. fabricated the device. S.A., K.P., S.L., M.-J.L. and M.J. measured the devices and analyzed the data, with help from J.S., M.B.A.J., V.K. and K.P. and J.P. performed nanowire synthesis. V.K., Z.B.S., J.H.B. and M.B.A.J performed the numerical simulations. M.J. and J.S. supervised the experiments presented in this paper. S.A., J.S. and M.J. prepared the paper with input from all authors.

**Notes**

The authors declare no competing financial interest.

## Table of Contents (TOC)

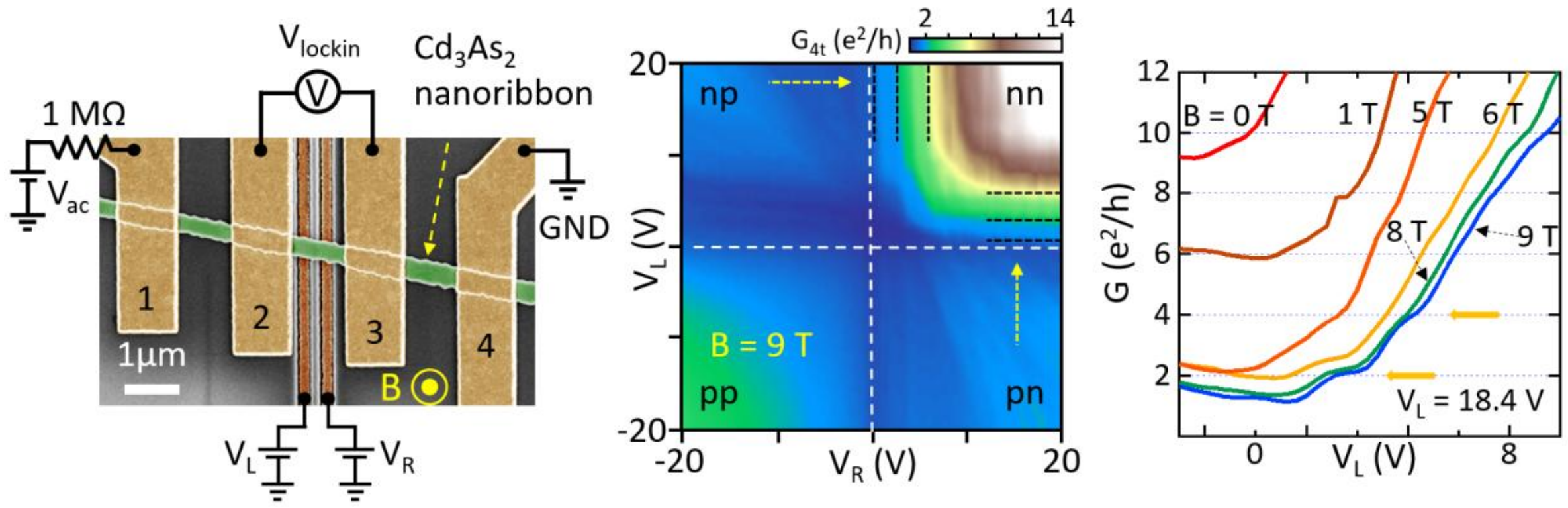

# Supporting Information:

# Quantized Conductance through Surface States in High Quality Three-Dimensional Dirac Semimetal $Cd_3As_2$ Nanowire/Nanoribbon p-n Junctions

Sungjin An,[1,¶] Zhuo Bin Siu,[2,¶] Vardan Kaladzhyan,[3¶] Jens H. Bardarson,[3] Sunghun Lee,[4] Myoung-Jae Lee,[1] Kidong Park,[1] Jeunghee Park,[5] Mansoor B. A. Jalil,[2] Jungpil Seo,[6,*] and Minkyung Jung[1,7,8*]

[1] DGIST Research Institute, DGIST, Daegu 42988, Korea

[2] Electrical and Computer Engineering, National University of Singapore, Singapore 117576, Republic of Singapore

[3] Department of Physics, KTH Royal Institute of Technology, Stockholm, SE-106 91 Sweden

[4] Industrial AX Innovation Research Institute and Sensor Technology Support Center, DGIST, Daegu 42988, Korea

[5] Department of Advanced Materials Chemistry, Korea University, Sejong 30019, Korea

[6] Department of Physics and Chemistry, DGIST, Daegu 42988, Korea

[7] Department of Interdisciplinary Engineering, DGIST, Daegu 42988, Korea

[8] Institute of next generation semiconductor technology (INST), DGIST, Daegu 42988, Korea

¶ *S.A, Z.B.S and V.K contributed equally to this work.*

[*]Corresponding authors. Email: jseo@dgist.ac.kr, minkyung.jung@dgist.ac.kr

## Supplement Note. S1: Contact resistance in Device A

We extracted the contact resistance of Device A by comparing two-terminal and four-terminal resistance maps measured as a function of $V_L$ and $V_R$. The two-terminal configuration includes both the intrinsic channel resistance and the contact resistance at the metal-nanowire interfaces, whereas the four-terminal configuration exhibits the intrinsic resistance by eliminating voltage drops at the contacts. Figures S1(a) and (b) show the measured two-terminal resistance ($R_{2t}$) and four-terminal resistance ($R_{4t}$) maps, respectively. Both maps exhibit similar overall features as a function of gate voltages, indicating consistent electrostatic control of the device. The contact resistance $R_C$ is extracted using $R_C = (R_{2t} - R_{4t})/2$, as shown in Fig. S1(c). The extracted $R_C$ ranges from approximately 2 to 5 kΩ, depending on the gate voltages.

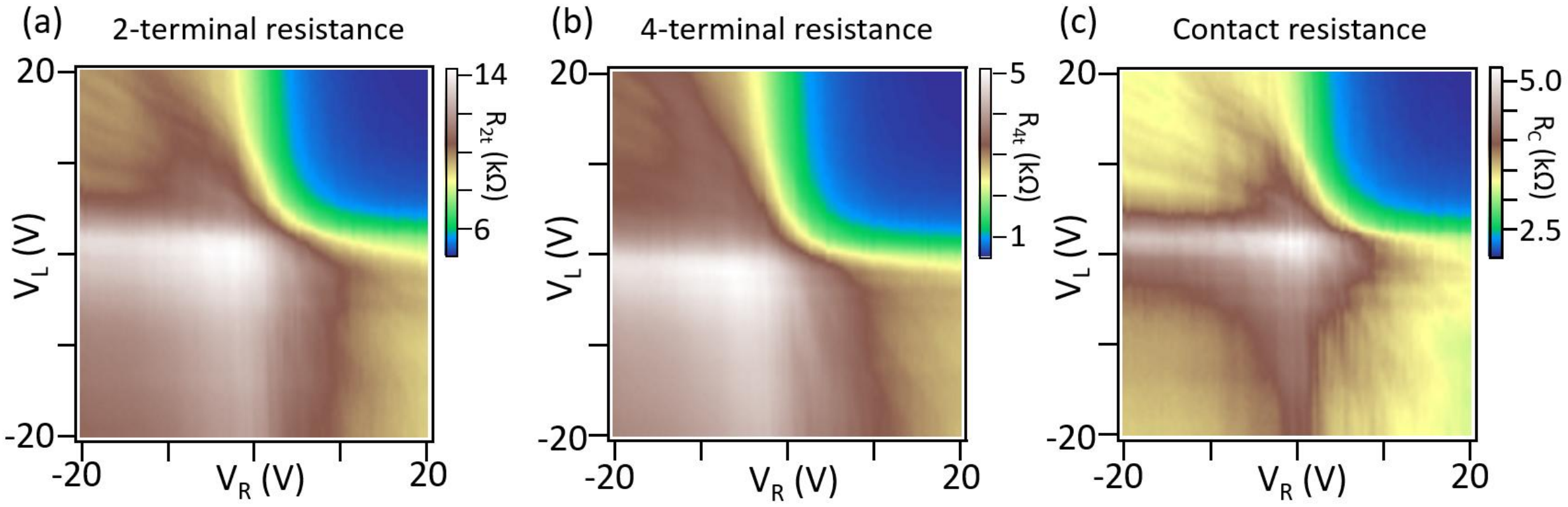

**Figure S1.** (a) Two-terminal resistance ($R_{2t}$) map and (b) four-terminal resistance ($R_{4t}$) map measured as a function of $V_L$ and $V_R$. (c) Contact resistance ($R_C$) extracted from the difference between the two configurations using $R_C = (R_{2t} - R_{4t})/2$.

**Supplement Note. S2: Modelling and numerical simulations**

To model the electronic and transport properties of the $Cd_3As_2$ nanowire, we employ a low-energy effective Hamiltonian describing a three-dimensional Dirac (Weyl) semimetal, following Refs. [18, 27, 28 in the main text]. This model captures the essential band structure near the Dirac nodes and is widely used to describe $Cd_3As_2$ nanostructures.

The Hamiltonian is given by:

$$H(k) = \varepsilon(k)\sigma_0\tau_0 + M(K)\sigma_z\tau_0 + Ak_x\sigma_x\tau_z + Ak_y\sigma_y\tau_z$$

where

$$\varepsilon(k) \equiv C_1k_z^2 + C_2(k_x^2 + k_y^2)$$

$$M(k) \equiv M_0 + |M_3| + \frac{M_1}{2|M_3|}k_z^2 + M_2(k_x^2 + k_y^2)$$

The model parameters are taken from Ref. [27]:

$C_1$ = –0.3 eV · nm$^2$, $C_2$ = –0.16 eV · nm$^2$, $A$ = 0.275 eV · nm,

$M_0$ = –0.06 eV, $M_1$ = 0.96 eV · nm$^2$, $M_2$ = 0.18 eV · nm$^2$, $M_3$ = 0.05 eV.

The momenta ($k_x, k_y, k_z$) are along the principal axes of the crystal. In order to calculate different properties we discretize it on a cubic lattice with lattice constant $a$ in the following way:

$$k_x \to \frac{1}{a}sink_xa,\ k_y \to \frac{1}{a}sink_ya,\ k_x^2 \to \frac{2}{a^2}(1 - cosk_xa),\ k_y^2 \to \frac{2}{a^2}(1 - cosk_xa),$$

$$k_z^2 \to \frac{2}{a^2}(1 - cosk_za).$$

We take the lattice spacing $a$ = 1.265 nm and consider a wire of dimension $L_x$ = 120 nm, $L_y$ = $L_z$ = 30 nm, with an applied magnetic field along the $z$-axis. For computational efficiency, the simulated nanowire is thinner than the experimental device, while preserving the essential confinement physics. We take into account both the orbital and the Zeeman effects of the magnetic field, using $\mu_B$ = 0.058 meV/T, $g_s$ = 18.6 and $g_p$ = 2. The conductance is obtained within the Landauer formalism for quantum transport. The numerical simulations shown in the main text were performed using the Kwant code [C. W. Groth, M. Wimmer, A. R. Akhmerov, X. Waintal, Kwant: a software package for quantum transport, *New J. Phys*. 16, 063065 (2014).]